\documentclass[journal=jacsat,manuscript=article]{achemso}

\usepackage[version=3]{mhchem}
\usepackage{graphicx}%
\usepackage{multirow}%
\usepackage{amsmath,amssymb,amsfonts}%
\usepackage{amsthm}%
\usepackage{mathrsfs}%
\usepackage[title]{appendix}%
\usepackage{xcolor}%
\usepackage{textcomp}%
\usepackage{manyfoot}%
\usepackage{booktabs}%
\usepackage{algorithm}%
\usepackage{algorithmicx}%
\usepackage{algpseudocode}%
\usepackage{listings}%

\theoremstyle{note}

\newtheorem{lem}{Lemma}

\theoremstyle{definition}

\theoremstyle{remark}

\newtheorem{thm*}{Theorem}

\makeatletter
\expandafter\newcommand\csname thethm*default\endcsname{\thethm*}
\newcommand{\thmstarnum}[1]{\expandafter\gdef\csname thethm*\endcsname{#1*}}
\thmstarnum{\thethm}
\expandafter\g@addto@macro\csname endthm*\endcsname{\thmstarnum{\thethm}}
\makeatother

\newtheorem{lem*}{Lemma}

\makeatletter
\expandafter\newcommand\csname thelem*default\endcsname{\thelem*}
\newcommand{\lemstarnum}[1]{\expandafter\gdef\csname thelem*\endcsname{#1*}}
\lemstarnum{\thelem}
\expandafter\g@addto@macro\csname endlem*\endcsname{\lemstarnum{\thelem}}
\makeatother

\newcommand{\sgn}{\operatorname{sgn}}

\newcommand{\rme}{\mathrm{e}}
\newcommand{\rmi}{\mathrm{i}}

\newcommand{\rmif}{\mathrm{if}}

\def\eqref#1{\textup{(\ref{#1})}}  
\newcommand{\eref}[1]{Eq.~\textup{(\ref{#1})}}

\newcommand{\tref}[1]{Table~\ref{#1}}

\newcommand{\cref}[1]{Conjecture~\ref{#1}}
\newcommand{\Cref}[1]{Conjecture~\ref{#1}}

\newcommand{\rcite}[1]{Ref.~\cite{#1}}

\title{Quantum molecular docking with quantum-inspired algorithm}

\author{Yunting Li}
\affiliation{Central Research Institute, Huawei Technologies, Shenzhen, 518129, China}
\alsoaffiliation{State Key Laboratory of Surface Physics and Department of Physics, Fudan University, Shanghai 200433, China}

\author{Xiaopeng Cui}
\affiliation{Central Research Institute, Huawei Technologies, Shenzhen, 518129, China}

\author{Zhaoping Xiong}
\affiliation{Laboratory of Health Intelligence, Huawei Cloud Computing Technologies Co., Ltd, Guizhou, 550025, China}

\author{Bowen Liu}
\affiliation{Central Research Institute, Huawei Technologies, Shenzhen, 518129, China}

\author{Bi-Ying Wang}
\affiliation{Central Research Institute, Huawei Technologies, Shenzhen, 518129, China}

\author{Runqiu Shu}
\affiliation{Central Research Institute, Huawei Technologies, Shenzhen, 518129, China}

\author{Nan Qiao}
\email{qiaonan3@huawei.com}
\affiliation{Laboratory of Health Intelligence, Huawei Cloud Computing Technologies Co., Ltd, Guizhou, 550025, China}

\author{Man-Hong Yung}
\email{yung.manhong@huawei.com}
\affiliation{Central Research Institute, Huawei Technologies, Shenzhen, 518129, China}
\alsoaffiliation{Shenzhen Institute for Quantum Science and Engineering, Southern University of Science and Technology, Shenzhen, 518055, China}
\alsoaffiliation{International Quantum Academy, Shenzhen, 518048, China}
\alsoaffiliation{Guangdong Provincial Key Laboratory of Quantum Science and Engineering, Southern University of Science and Technology, Shenzhen, 518055, China}
\alsoaffiliation{Shenzhen Key Laboratory of Quantum Science and Engineering, Southern University of Science and Technology, Shenzhen, 518055, China}

\begin{document}

\begin{abstract}
Molecular docking (MD) is a crucial task in drug design, which predicts the position, orientation, and conformation of the ligand when bound to a target protein. It can be interpreted as a combinatorial optimization problem, where quantum annealing (QA) has shown promising advantage for solving combinatorial optimization. In this work, we propose a novel quantum molecular docking (QMD) approach based on QA-inspired algorithm. We construct two binary encoding methods to efficiently discretize the degrees of freedom with exponentially reduced number of bits and propose a smoothing filter to rescale the rugged objective function. We propose a new quantum-inspired algorithm, hopscotch simulated bifurcation (hSB), showing great advantage in optimizing over extremely rugged energy landscapes. This hSB can be applied to any formulation of objective function under binary variables. An adaptive local continuous search is also introduced for further optimization of the discretized solution from hSB. Concerning the stability of docking, we propose a perturbation detection method to help ranking the candidate poses. We demonstrate our approach on a typical dataset. QMD has shown advantages over the search-based Autodock Vina and the deep-learning DIFFDOCK in both re-docking and self-docking scenarios. These results indicate that quantum-inspired algorithms can be applied to solve practical problems in the drug discovery even before quantum hardware become mature.
\end{abstract}

\maketitle

\section{Introduction}

Efficient drug discovery is of significant interest in pharmaceutical science. In the last few decades, structure-based drug design with computer-assisted methods \cite{Leelananda2016,dosSantos2018,Vikram2020} has accelerated the discovery of drug candidates. 
\emph{Molecular docking} (MD) is a crucial task in drug design, which predicts the position, orientation, and conformation of the ligand when bound to a target protein \cite{KUNTZ1982269,Daniel1996MD,Huang2006MD}. There are two common scenarios related to the configuration of the ligand in MD \cite{TANKBind2022}: \emph{re-docking}, where the ligand is regarded as rigid body with given and fixed conformation; \emph{self-docking}, where all degrees of freedom of the ligand (position, orientation, and conformation) are variables. In addition, according to the range of docking, there are also two scenarios \cite{TANKBind2022}: \emph{pocket docking}, which is practical and widely-concerned in real experiments with a given possible docking pocket of the protein; \emph{blind docking}, where no pocket information is given and the ligand is allowed docking anywhere on the protein.

Traditional approaches are based on systematic or stochastic search algorithms to sample in the search space and rank the predictions according to a score function \cite{STANZIONE2021273}. In general, the score function is usually related to potential energy and can be physics-based, empirical, knowledge-based or machine learning-based \cite{Li2014,Li2019}. Many search-based softwares \cite{C6CP01555G,Pagadala2017} are available: AutoDock \cite{autodock1998,autodockvina2010,autodockvina2021}, GLIDE \cite{GLIDE2004}, SMINA \cite{SMINA2013}, and GNINA \cite{GNINA2021} et. al.. However, the rugged energy landscape of MD problem is difficult for classical algorithms to find the global minimum.
Another method based on subgraph matching has emerged in recent years \cite{Banchi2020}, which represents both the ligand and the protein pocket as a subgraph with pharmacophores as vertexes. According to the interaction energy between the ligand and the protein pharmacophores, a binding interaction graph is constructed and the MD problem is reduced to finding the maximum weighted clique in the graph. 
A recent breakthrough, DIFFDOCK \cite{corso2023diffdock}, is based on a diffusion generative model, which defines a diffusion process transforming the data distribution in a tractable prior and learn the score function of the evolving distribution by machine learning. This method based on blind docking has an impressive 30\% success rate, i.e., percentage of root-mean-square distance (RMSD) below 2Å.


In recent years, quantum computing has emerged as a promising tool to accelerate the computation and reduce resources by exploiting quantum mechanical systems \cite{Han2002,Ewin2019,Rss2020,Arrazola2020quantuminspired,Oshiyama2022,Kowalsky_2022,huang2022benchmarking,Zeng2024}.
Many attempts of structure-based drug design with quantum computing have appeared \cite{Y2018IBM,Carlos2021,ZINNER20211680,Blunt2022,Mato2022,pandey2022multibody,yan2023quantum}. In the conformational generation problem, Mato et al. have applied \emph{quantum annealing} (QA) \cite{Mato2022}. Several studies have focused on the protein folding problem based on the Quantum Approximate Optimization Algorithm (QAOA) \cite{Robert2021,Babbush2014,Perdomo2012,fingerhuth2018quantum,babej2018coarsegrained}.
In addition, Pandey et al. have realized a simplized multibody docking problem on a quantum annealer \cite{pandey2022multibody}.
To address molecular docking problem, we explore the effectiveness of quantum-inspired algorithms \cite{Zhang2011,Rss2020,Arrazola2020quantuminspired,liu2020} for support, which are expected to tunnel through "high but narrow" barriers in energy landscapes. Specifically, we employ the simulated bifurcation (SB) algorithm, which is a purely quantum-inspired algorithm based on quantum adiabatic optimization using nonlinear oscillators \cite{Hayato2016,Hayato2019,Hayato2019JPS,Hayato2021,Kanao2022,Dmitry2021}. SB is driven by Hamiltonian dynamics evolution, and allows for the simultaneous updating of variables, resulting in great parallelizability and accelerated combinatorial optimization. SB also shows significant potential for finding the global minimum. Recently, two SB variants, namely \emph{ballistic SB} (bSB) and \emph{discrete SB} (dSB), have been proposed for further enhancement~\cite{Hayato2021}.


In this work, we propose an efficient \emph{quantum molecular docking} (QMD) approach to solve MD problem with quantum-inspired algorithms. 
The Gray code method is applied to discrete and encode the degrees of freedom of ligand into binary variables. We also construct a phase encoding method to efficiently encode angular variables.
We determine the interaction energy function as the objective function in MD optimization problem, and propose a smoothing filter to rescale the rugged objective function, which significantly benefits further optimization.
In addition, we propose a new SB variant, \emph{hopscotch SB} (hSB), which has great advantage in optimizing over rugged energy landscapes. This hSB can be applied to any formulation of objective function under binary variables.
An adaptive local continuous search is also introduced for further optimization of the discretized solution from hSB. Concerning the stability of docking, we propose a perturbation detection method to help ranking the candidate poses.
We demonstrate the performance of our QMD approach on the same testing dataset as DIFFDOCK with respect to RMSD and success rate metrics. 
In pocket re-docking scenario, QMD increases the top-1 success rate of the root-mean-square distance (RMSD) less than 2Å by about 6\% compared with DIFFDOCK.
In pocket self-docking scenario, QMD with perturbation detection increases the top-1 success rate by about 2\% compared with Autodock Vina, and increases the top-5 success rate of RMSD less than 5Å by about 6\% compared with DIFFDOCK.

\section{Background}

\subsection{Molecular docking problem}\label{sec:MD}

Molecular docking problem aims to predict the position, orientation and conformation of a ligand (usually referred to as pose) when bound to a target protein. A score function is introduced to estimate the correctness of an optimized pose and rank several optimized samples.

In general, the target protein is rigid and the structure is given. The self-docking allows the degrees of freedom of the ligand including three translations $T_x, T_y, T_z$ in Cartesian coordinate system, three rotation angles $\phi_x, \phi_y, \phi_z$ around the ligand itself, and $M$ torsion angles $\theta_1, \dots, \theta_M$ along $M$ rotatable bonds of the ligand. The re-docking refers to only consider the first two rigid transformations (translation and rotation) with a fixed conformation given as input.

Suppose the position of an atom $i$ in the ligand is denoted by $\vec{r}_i$. Without loss of generality, the ligand is updated in the sequence of rotations, translations and torsions. The translations and rotations update the atom position as follows:
\begin{align}
\vec{r}'_i = R(\phi_x, \phi_y, \phi_z) (\vec{r}_i - \vec{r}_0) + \vec{r}_0 + \vec{T},
\end{align}
where $\vec{r}_0$ is the geometric center of the current ligand, $\vec{T}=(T_x, T_y, T_z)^T$ is a vector giving the translation of the origin, and $R(\phi_x, \phi_y, \phi_z) = R_z(\phi_z)R_y(\phi_y)R_x(\phi_x)$ is the rotation matrix obtained from three basic rotation matrices of angles $\phi_x, \phi_y, \phi_z$ around $x, y, z$ axis. 
Next the torsions update the atom position $\vec{r}'_i$ as follows:
\begin{align}
\vec{r}''_i = \mathbf{R}_M \circ \cdots \circ \mathbf{R}_2 \circ \mathbf{R}_1 (\vec{r}'_i),
\end{align}
where $\mathbf{R}_i, i=1, \dots, M$ is the torsion update of angle $\theta_i$ along the $i$th rotatable bond. The details of torsion update are elaborated in Sec.~S1 in the Supplemental Material \cite{supp}.

The molecular docking process can be viewed as an optimization problem: minimizing the interaction energy over the degrees of freedom of the ligand to generate a stable complex of ligand and protein.
The interaction energy between the ligand and protein includes the Lennard-Jones potential energy and the electrostatic potential energy, which is usually regarded as the objective function and the score function.

To be specific, the Lennard-Jones potential models soft repulsive and attractive (van der Waals) interactions for electronically neutral atoms or molecules. The commonly used expression reads \cite{autodockvina2010, autodockvina2021}
\begin{align}\label{eq:V_LJ}
V_{\mathrm{LJ}} = \sum_{i,j} V_{\mathrm{LJ}}^{ij} = \sum_{i,j} \varepsilon_{ij} (\frac{R_{ij}}{r_{ij}})^{12} - 2  \varepsilon_{ij} (\frac{R_{ij}}{r_{ij}})^{6},
\end{align}
where $i,j$ represent the atoms in the ligand and the protein respectively, $R_{ij}$ is the sum of radius of atom $i$ and $j$, $r_{ij}$ is the distance between atom $i$ and $j$, and $\varepsilon_{ij}$ is the depth of the potential well of atom $i$ and $j$.

The electrostatic potential energy results from conservative Coulomb forces and is associated with the configuration of a particular set of point charges within a defined system. The commonly used expression reads \cite{autodockvina2010, autodockvina2021}
\begin{align}\label{eq:V_e}
V_e = \sum_{i,j} V_e^{ij} = \sum_{i,j} \frac{q_i q_j}{f(r_{ij}) r_{ij}},
\end{align}
where $q_i,q_j$ are the charges of atom $i$ and $j$ respectively, $r_{ij}$ is the distance between atom $i$ and $j$, and $f(r_{ij})$ is a correction function of $r_{ij}$. 

Therefore, the total interaction energy is a weighted intermolecular pair potential reads \cite{autodockvina2010, autodockvina2021}
\begin{align}\label{eq:interaction energy}
V = w_1 V_{\mathrm{LJ}} + w_2 V_e,
\end{align}
where $w_1, w_2$ are the weights for two potential energy. 
The degrees of freedom of the ligand can affect positions of each atom resulting in the change of $r_{ij}$ terms. The optimization problem is to minimize the total interaction energy $V$ as the objective function. The details of parameters, the correction function and other techniques are elaborated in Sec.~S2 in the Supplemental Material \cite{supp}.

\subsection{Quantum-inspired algorithms}

The original simulated bifurcation (SB) \cite{Hayato2016,Hayato2019,Hayato2019JPS,Hayato2021} is a heuristic technique for accelerating combinatorial optimization in the Ising formulation. 
A network of quantum-mechanical nonlinear oscillators \emph{Kerr-nonlinear parametric oscillators} (KPOs) is introduced in the quantum Hamiltonian \cite{Hayato2016,Hayato2019}.
Driven by the derived classical Hamiltonian dynamics evolution from Hamilton-Jacobi equation, the equations of motion of position $x_i$ and momentum $y_i$ of the $i$th particle can be obtained. All positions and momentums are randomly set around zero at the initial time.
During the evolution, the positions $x_i$ of each particle oscillate and then move towards values $\pm 1$, exhibiting as bifurcations.
Finally, the $\sgn(x_i)$ gives the solution of the corresponding spin value $s_i$ in the Ising formulation.

From the intuition of binary spin values, a variant bSB is introduced \cite{Hayato2021} with inelastic walls set at $x_i=\pm 1$.
Moreover, the discrete version of bSB, namely, dSB is introduced \cite{Hayato2021} to suppress analog deviation by discretizing $x_i$ with $\sgn(x_i)$ when updating the derivative of the momentum $y_i$. 

Fortunately, recent study has shown that dSB can be potentially generalized to higher-order unconstrained binary objective (HUBO) formulation and demonstrated the advantage in a three-order task \cite{Kanao2022}. By replacing the Ising energy term with a general higher-order objective function $E(\vec{x})$, we have the general equations of motion as follows:
\begin{equation}
\begin{aligned}\label{eq:dSB HUBO}
\dot{x}_i &= \frac{\partial H_{\mathrm{dSB}}}{\partial y_i} = a_0 y_i \\
\dot{y}_i &= -\frac{\partial H_{\mathrm{dSB}}}{\partial x_i} = \left[a_0 -a(t)\right] x_i + c_0 \left.\frac{\partial E(\vec{x})}{\partial x_i}\right|_{\vec{x}=\sgn(\vec{x})},
\end{aligned}
\end{equation}
where $\vec{x}$ is the vector of positions of all spins, $x_i, y_i$ are the positions and momentums of a particle corresponding to the $i$th spin, $a(t)$ is a control parameter increased from zero to $a_0$, and $a_0, c_0$ are positive constants.
The Hamiltonian $H_{\mathrm{dSB}}$ reads 
\begin{equation}
H_{\mathrm{dSB}} = \frac{a_0}{2} \sum_i^N y_i^2 + V_{\mathrm{dSB}},
\end{equation}
and if $ |x_i| \le 1 $ for all $x_i$, the potential $V_{\mathrm{dSB}}$ reads
\begin{equation}
V_{\mathrm{dSB}} =
\frac{a_0-a(t)}{2} \sum_i^N x_i^2  - c_0 E(\vec{x}),
\end{equation}
otherwise it reads infinity.

The algorithmic advantages of SB are that the equation of motion is easy to implement with only product-sum operation given the derivatives; the variables can be updated simultaneously for parallel acceleration; and SB does not contain probabilistic processes, and therefore can be easily hardwired \cite{Hayato2019}.

Since SB algorithms require binary variables, encoding methods, including binary encoding, Gray encoding \cite{graycode1986} and one-hot encoding \cite{onehot2015}, are necessary to encode general variables into binary variables in general tasks. A suitable choice of encoding methods may also improve the efficiency and accuracy for the particular task.

\section{Methods}

\subsection{Overview}\label{sec:framework}

\begin{figure*}[ht]
\centering
\includegraphics[scale=0.38]{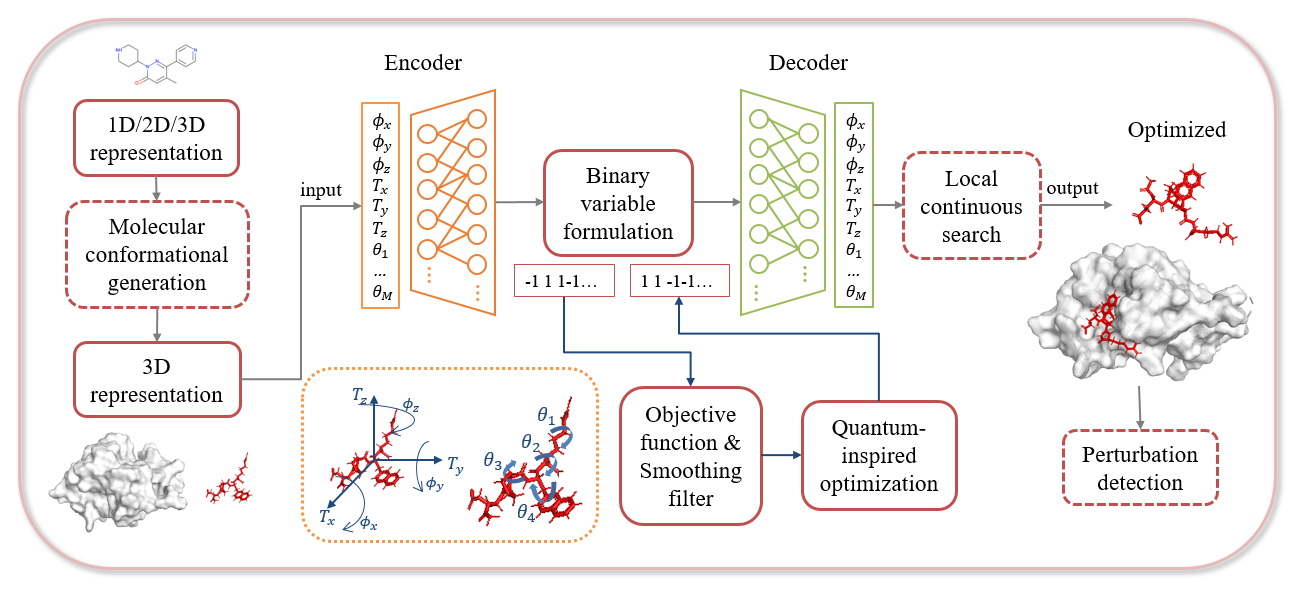}
\caption{The framework of Quantum-inspired Molecular Docking. The inputs are the 3D representations of the ligand and protein which can be generated by an optional module (molecular conformational generation). The degrees of freedom are discretized and encoded to the binary variable formulation. The objective function is of binary variables based on potential energy and then transformed through a smoothing filter. After optimized by quantum-inspired algorithms, the optimized binary result is decoded to discrete values of the degrees of freedom. The local continuous search is an optional module for further optimization in the continuous domain. The optimized continuous result is output to recover the optimized pose of the ligand. The perturbation detection is an optional module for helping rank several optimized poses by virtue of the stability.}
\label{fig:framework}
\end{figure*}

The framework in Figure~\ref{fig:framework} shows a general process of quantum-inspired molecular docking.
The inputs are the 3D representations of the ligand and the protein. If only an 1D/2D representation or a rough 3D representation is available, we introduce an optional module, molecular conformational generation, to generate the 3D representation based on potential energy \cite{rdkit} or molecular volume \cite{Mato2022}.
The degrees of freedom of the ligand including translations $T_x, T_y, T_z$, rotations $\phi_x, \phi_y, \phi_z$ and torsions $\theta_1, \dots, \theta_M$ are discretized and encoded to binary variables. The objective function is of binary variables based on potential energy and then transformed through a smoothing filter. After optimized by quantum-inspired algorithms in the binary domain, the solution is then decoded to the corresponding discrete values of the degrees of freedom. For further optimization in the continuous domain, an optimized continuous solution is given by a local continuous search. Finally, the output recovers the optimized pose of the ligand. 

After we obtain several optimized poses, the best pose is ranked by the score function.
Considering the stability of the crystallization in docking experiments, the perturbation detection is an optional module to better rank the candidate poses.

\subsection{Discretization formulation}

The degrees of freedom can be discretized into uniformly discrete values by encoding methods.
Here we generalize the gray code method with coefficients to encode any range of uniformly discrete values. We also adapt the phase encoding method to encode trigonometric functions.
If the degree of freedom corresponds to $n$ binary variables, then it has ${d=2^n}$ uniformly discrete values.
The gray code is suitable for all three kinds of variables (translation, rotation and torsion), while phase encoding is suitable for encoding angle variables (rotation and torsion).

\subsubsection{Gray code method}

In gray code method \cite{graycode1986}, the bit strings of two adjacent integers differ in only one bit.
Given $n$ binary variables $\mathbf{s} = (s_0 , \dots, s_{n-1}) \in \{-1,1\}^n$, the correspondence to ${d=2^n}$ integers ${\{-d/2+1,-d/2+2, \dots, d/2\}}$ is as follows:
\begin{align}
D(\mathbf{s}) = \frac{1}{2} \left(\sum_{j=0}^{n-1} 2^{n-j-1} \prod_{i=0}^j s_i + 1 \right).
\end{align}

Suppose the step sizes of translation, rotation and torsion are $\Delta T, \Delta \phi, \Delta \theta$ respectively. Then the discrete values of each degree of freedom are $ \Delta T D(\mathbf{s}), \Delta \phi D(\mathbf{s}), \Delta \theta D(\mathbf{s})$. To ensure sufficient sampling, $\Delta T$ is at most 0.5Å and $\Delta \phi, \Delta \theta$ are about $5^\circ$.
Here we set $\Delta T=T/d$ where $T$ is the translation range, and $\Delta \phi = \Delta \theta = 2\pi/d$. In general, the value of $d$ can be different for translation, rotation and torsion depending on the needs.

\subsubsection{Phase encoding method}

In phase encoding method \cite{PE2023}, the correspondence between the phase $\rme^{\rmi \psi}$ of an angle $\psi \in \{\psi_k\}$ and a polynomial $p_n(\mathbf{s})$ is established.
The detail is shown in the following lemma from \rcite{PE2023}.

\begin{lem}
Suppose $n$ is a positive integer and $\{\psi_k\}_{k=0}^{2^n-1}$ are phases uniformly distributed on the circle. 
Then we can construct a polynomial $p_n(\mathbf{s})$ with $2^{n-1}$ terms such that $\{p_n(\mathbf{s})|\mathbf{s}\in \{\pm 1\}^n\}=\{\rme^{\rmi \psi_k}\}_{k=0}^{2^n-1}$. 
\end{lem}

Since the angles only appear in the rotation operator in the form of $\sin\psi, \cos\psi$, then we propose the efficient polynomial expressions of $\sin\psi, \cos\psi$ from the polynomial expression $p_n(\mathbf{s})$ of $\rme^{\rmi \psi}$ read
\begin{align}
\sin\psi = \Im p_n(\mathbf{s}), \quad \cos\psi = \Re p_n(\mathbf{s}).
\end{align}
In addition, we can also obtain the efficient polynomial expression of $\psi$ as follows:
\begin{align}
\psi = -\rmi \ln p_n(\mathbf{s}).
\end{align}

If $n=3$ for example, we have $\psi_k =\frac{2\pi}{2^n} k = \frac{\pi}{4} k$ and the corresponding polynomial reads
\begin{equation}
p_3(\mathbf{s})  = c_0 s_0 + c_1 s_1 + c_2 s_2 + c_3 s_0 s_1 s_2, 
\end{equation}
where $\mathbf{s}=(s_0, s_1, s_2) \in \{\pm 1\}^3$ and the coefficients are
\begin{equation}
\begin{pmatrix}
c_0 \\ c_1 \\ c_2 \\ c_3
\end{pmatrix}
=\frac{1}{4}
\begin{pmatrix}
1+(\sqrt{2}-1)\rmi \\ 1-(\sqrt{2}+1)\rmi \\ (1+\sqrt{2})+\rmi \\ (1-\sqrt{2})+\rmi
\end{pmatrix}.
\end{equation}
Then the polynomial expressions of $\sin\psi, \cos\psi$ can be obtained from straightforward calculations.
Similarly, for any value of $n$, we can obtain the polynomial expression of discrete phases according to \rcite{PE2023}.

Therefore, replacing the denotation $\psi$ with the concerning rotation angles $\phi_x, \phi_y, \phi_z$ and torsion angles $\theta_1, \dots, \theta_M$, we obtain the encoding correspondence between the $d$ discrete values and the $n$ binary variables~$\mathbf{s}$. The step sizes of rotation angles and torsion angles are $\Delta \phi = \Delta \theta = 2\pi/d$. In general, the value of $d$ can also be different for rotation and torsion depending on the needs.

\begin{figure}[t]
\centering
\includegraphics[scale=0.4]{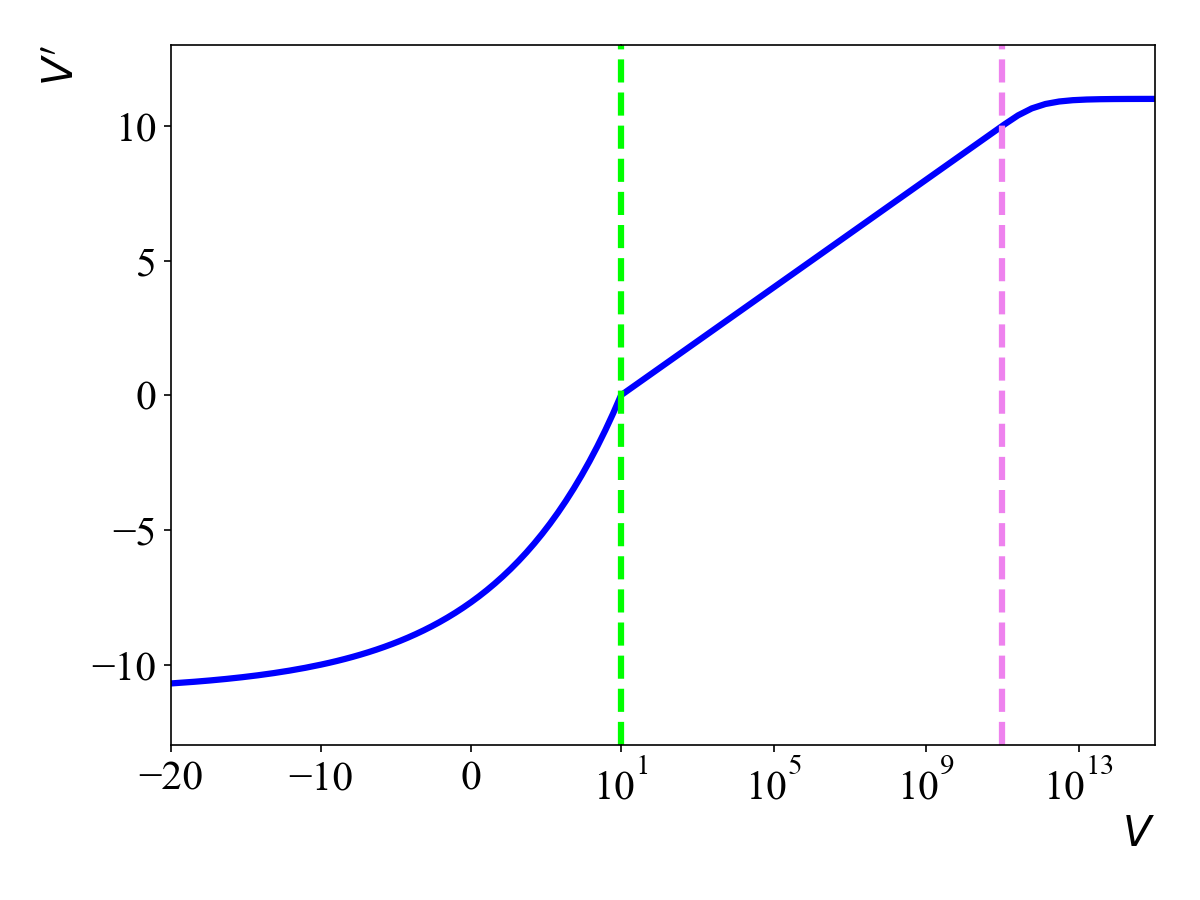}
\caption{The effect of the smoothing filter. We set $V_0=10$, $H_0 = -10$, $\Delta V=20$ and thus $x_0=10$. The two dashed vertical lines separate the domain into three areas corresponding to the three cases in \eref{eq:smoothing} respectively.}
\label{fig:smoothing}
\end{figure}

\subsection{Smoothing filter}

To further facilitate the optimization of rough objective function, we propose the smoothing filter technique to compress the huge peak and expand the sharp minimum. Since in MD the objective function is inversely proportional to the 12th power of the distance between two atoms in \eref{eq:interaction energy}, the close atom pair will cause huge value of objective function, which is not suitable for traditional optimization algorithms. 

The smoothing filter is realized by a map function.
We map the original objective function $V$ defined in \eref{eq:interaction energy} to a bounded range $(-V_0-1, V_0+1)$, $V_0 > 0$. The map function is designed to maintain the monotonicity, flatten the objective function around a given low enough value, and greatly compress the huge peak of the objective function. The form of map functions is not unique and can be constructed differently according to the specific problem. 

Here we construct a map function as follows:
\begin{align}\label{eq:smoothing}
V' =
\begin{cases}
\exp[ \frac{\log (V_0+1)}{x_0-H_0}(V-H_0)] - (V_0+1), \quad \rmif \ V \le x_0, \\
\log V - \log x_0 + x_0, \quad \rmif \ x_0 < V \le 10^{\log x_0+V_0}, \\
\tanh(\log V - \log x_0 - V_0) + V_0, \quad \rmif \ V > 10^{\log x_0+V_0},
\end{cases}
\end{align}
where $H_0$ is the given low enough value with the similar magnitude of the minimum, $x_0 = {\max \{H_0 + \Delta V, 1\}}$ is the upper bound for flattening, and $\Delta V$ is the range of flattening. Once $V \ge x_0$, $V$ usually have huge values scaling as power of 10 due to the overlap of atom pairs. The effect of the smoothing filter is shown in Figure~\ref{fig:smoothing}. Here we set $V_0=10$, $H_0 = -10$, $\Delta V=20$ and $x_0=10$ for illustration, which is also suitable for molecular docking problem. The two dashed vertical lines separate the domain into three areas corresponding to the three cases in \eref{eq:smoothing} respectively.

Therefore, the objective function is transformed through the smoothing filter and then is used for optimization.

\subsection{Hopscotch simulated bifurcation}


In this work, we propose a new variant of SB, hSB, for optimizing rough objective function with sharp minimums and huge peaks. Additionally, hSB does not require the objective function to be in HUBO formulation and does not require the information of derivatives, and therefore is capable for any formulation under binary variables.

\begin{figure}[t]
\centering
\includegraphics[scale=0.25]{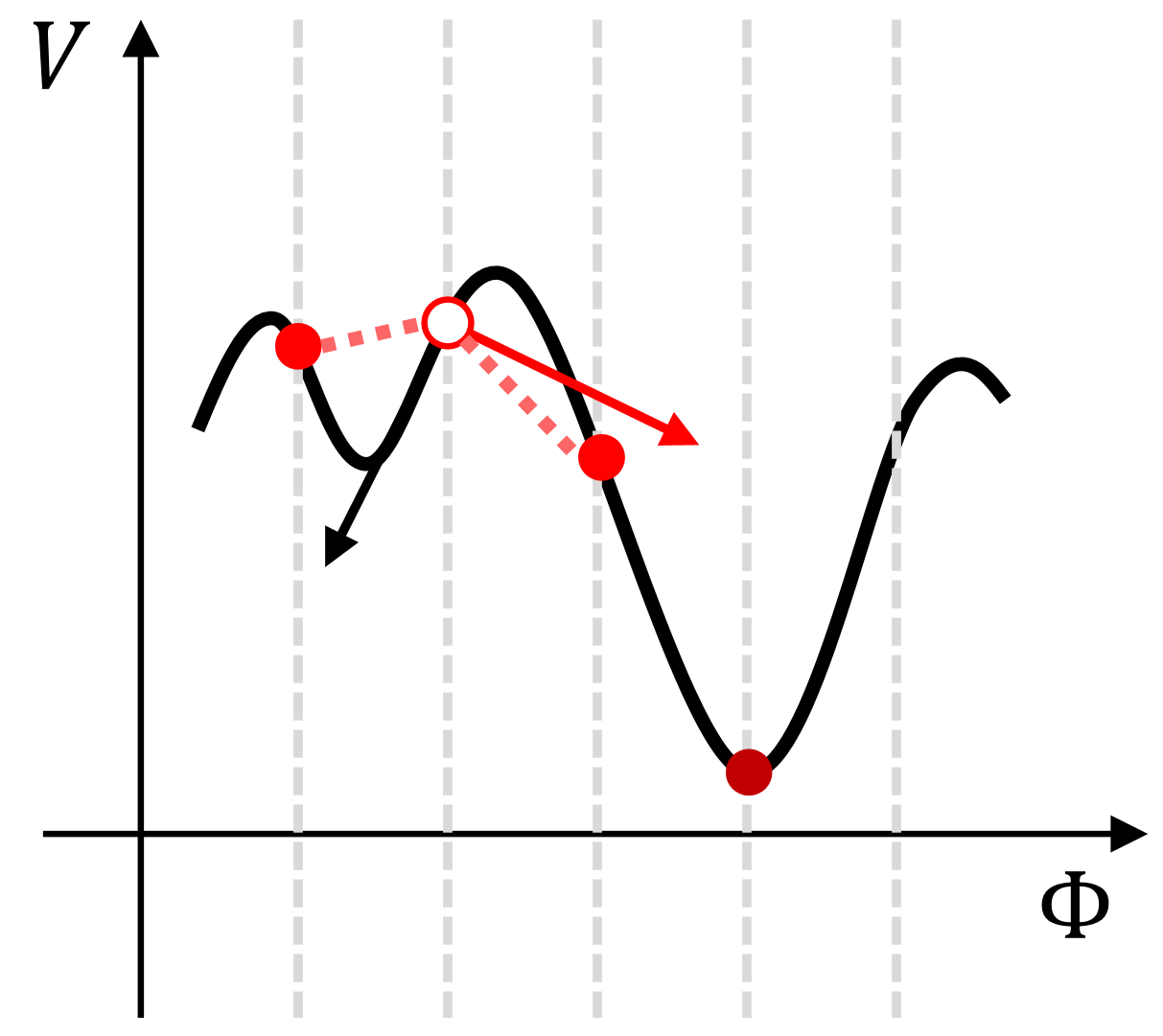}
\caption{The schematic diagram of the improvement in hSB. The variable space is discretized. The dark red dot (the most left) represents the global minimum. The black arrow indicates the derivative of the continuous objective function at the red circle position, while the red arrow indicates the discrete difference calculated from the nearest two red dots.}
\label{fig:hSB}
\end{figure}

Even though the partial derivative of $E(\vec{x})$ is calculated at $\vec{x}=\sgn(\vec{x})$ in dSB as shown in \eref{eq:dSB HUBO}, it still comes from the continuous domain. This may give wrong evolution directions since the derivative of a continuous position is not in correspondence with the one of discrete difference as shown in Figure~\ref{fig:hSB}.
The dark red dot (the most left) represents the global minimum. When evolving to the red circle position, the derivative of the continuous objective function will indicate the black arrow as the next evolution direction. However, the discrete difference calculated from the nearest two red dots will indicate the red arrow which is correct.
When there is a peak between two discrete positions, bSB and dSB may lead to wrong evolution path and miss the global minimum.

To avoid the misleading of information in the continuous domain, the evolution of hSB only depends on the discrete positions $\sgn(x_i)$. We replace the partial derivative $\left.\frac{\partial E(\vec{x})}{\partial x_i}\right|_{\vec{x}=\sgn(\vec{x})}$ with the difference of adjacent discrete positions as follows:
\begin{align}
-\frac{1}{2} \left( E[e_i(\sgn(\vec{x}))] - E[\sgn(\vec{x})] \right)\sgn(x_i) ,
\end{align}
where $e_i(\sgn(\vec{x}))$ is $\sgn(\vec{x})$ with the $i$th element flipped. 
Therefore, the influence from the non-discrete region is completely removed, and the optimization strictly depends on the objective function in discrete region. 

The hSB yields great improvements in optimizing rough objective functions and allows any formulation of the objective function.
Additionally, all variables can be simultaneously updated at each iteration step if parallel computation is used. The time cost of hSB can be theoretically reduced by the times of the number of variables if parallel computation is available.

\subsection{Local continuous search}

After the discrete optimization by quantum-inspired algorithm hSB, we propose a local continuous search to further optimize over the degrees of freedom of translations $T_x, T_y, T_z$, rotations $\phi_x, \phi_y, \phi_z$ and torsions $\theta_1, \dots, \theta_M$ in the continuous domain.

Here we consider the adaptive gradient descent (Adagrad) with adaptive learning rate. Denote by $f(\Theta)$ the objective function and $\Theta$ the variables. The evolution of Adagrad reads
\begin{equation}
\begin{aligned}
G_{t} &= G_{t-1} + \nabla f(\Theta), \\
\Theta_{t+1} &= \Theta_t - \frac{\alpha_t}{\sqrt{G_t + \epsilon}} \nabla f(\Theta),
\end{aligned}
\end{equation}
where $G_{t}$ is a vector of the squared sum of the past-time gradients with regards to each variable, $\epsilon$ is a correction term to avoid dividing by 0 and is generally insignificantly small ($10^{-8}$), and $\alpha_t$ is an adaptive learning rate to adjust the weight of $G_t$ for accelerating the optimization.

\subsection{Perturbation detection}

In general, the interaction energy is usually used as the score function to rank several optimized poses. However, the crystal pose is not exactly corresponding to the lowest-energy pose. In the docking experiments, the crystallization process is also related to the stability of the pose.
Therefore, we propose the perturbation detection to help rank the optimized poses.

For each optimized pose, we slightly perturb the pose for $M$ times by adding a small translation and rotation and record the corresponding interaction energy after each perturbation. We calculate the standard deviation of the $M$ interaction energy.
When ranking the optimized poses, we first filter the $k$ lowest energy within a small $\delta$ difference (e.g. $\delta=5\%$), then the best pose is the one with the smallest standard deviation within the $k$ poses. Note that if the lowest energy is greatly smaller than others, i.e., the difference is larger than $\delta$, then the best pose is still the one with the lowest energy.

\section{Results}\label{sec:results}

\subsection{Re-docking scenario}

In re-docking scenario, the target protein and the ligand are regarded as rigid bodies. Thus the ligand has three degrees of freedom of translation and three degrees of freedom of rotation. The ligand conformation is pre-generated and given as the input.
Here we compare our approach QMD with the ML-based method DIFFDOCK. During the inference process in DIFFDOCK, the ligand configuration can be fixed.
We evaluate on the same testing dataset of 363 protein-ligand pairs as DIFFDOCK \cite{corso2023diffdock}. For re-docking, the approaches receive the crystal structures of the protein and the ligand as inputs. 
10 initial poses of the ligand are randomly selected, and the corresponding predicted docking pose are ranked according to the score function in \eref{eq:interaction energy}. 
The implementation details of QMD and Autodock Vina are explained in Sec.~S3 in the Supplemental Material \cite{supp}.

\begin{table*}[t]
\caption{Re-docking comparison between QMD and DIFFDOCK. The top-1 RMSD is the RMSD of the highest ranked pose; the top-5 RMSD is the smallest RMSD among the 5 highest ranked poses. Shown is the percentage of predicted poses with RMSD $<2$Å and the mean RMSD for top-1 and top-5 respectively.}
\label{table:re-docking}
\centering
\renewcommand{\arraystretch}{1.5}
\begin{tabular}{@{}ccccc@{}}
\toprule
\multirow{2}*{\textbf{Method}} & \multicolumn{2}{c}{\textbf{Top-1 RMSD (Å)}} &  \multicolumn{2}{c}{\textbf{Top-5 RMSD (Å)}}  \\ 
~ & \%$<2$ & Mean & \%$<2$ & Mean  \\ \hline
DIFFDOCK (10) & $49.94\pm 0.78$ & $7.66\pm 0.21$ & $59.50\pm 1.40$ & $3.99\pm 0.05$  \\ \hline
QMD (10, no local) & $46.40\pm 2.65$ & $4.16\pm 0.18$ & $56.11\pm 2.63$ & $2.64\pm 0.11$ \\ \hline
QMD (10) & \pmb{$55.37\pm 1.68$} & \pmb{$3.59\pm 0.04$} & \pmb{$63.77\pm 2.04$} & \pmb{$2.37\pm 0.09$}  \\ \hline
\end{tabular}
\end{table*}

We consider the RMSD between the predicted docking pose and the ground-truth (crystal pose). 
The comparison of re-docking is shown in \tref{table:re-docking}.
The top-1 RMSD is the RMSD of the highest ranked pose; the top-5 RMSD is the smallest RMSD among the 5 highest ranked poses, which is a useful metric when multiple predictions are used for downstream verification and applications. The success rate (i.e., percentage of predicted poses with RMSD $<2$Å) and the mean RMSD for top-1 and top-5 are shown respectively.
The experiments are repeated for 3 times and the average success rate with standard deviation is shown. 
QMD improves the success rate about 5\% compared to DIFFDOCK, and the local continuous search in QMD can improve the success rate about 9\% based on the results from hSB.

\subsection{Self-docking scenario}

We conduct a comparative analysis between our approach with the search-based method Autodock Vina and DIFFDOCK in the pocket self-docking scenario, where the ligand has the degrees of freedom of translation, rotation and torsion.
Autodock Vina is implemented by the open Python package, ODDT (Open Drug Discovery Toolkit) \cite{ODDT2015}.
We evaluate on the same testing dataset of 363 protein-ligand pairs as DIFFDOCK \cite{corso2023diffdock}.
For self-docking, the approaches receive the crystal pose of the protein and the generated conformation of the ligand by UFF (Universal Force Field) optimization in RDKit. 
The pocket is of size $12\times12\times12$Å$^3$ centered at the crystal pose.
For each approach, 10 initial poses of the ligand are randomly selected, and the corresponding predicted docking pose are ranked according to the score function. In addition, we also consider the QMD with perturbation detection. The implementation details are explained in Sec.~S3 in the Supplemental Material \cite{supp}.

\begin{table*}[t]
\caption{Self-docking comparison between QMD and Autodock Vina. The top-1 RMSD is the RMSD of the highest ranked pose; the top-5 RMSD is the smallest RMSD among the 5 highest ranked poses. Shown is the percentage of predicted poses with RMSD $<2$Å and the mean RMSD for top-1 and top-5 respectively.
}
\label{table:self-docking}
\centering
\renewcommand{\arraystretch}{1.5}
\begin{tabular}{@{}ccccc@{}}
\toprule
\multirow{2}*{\textbf{Method}} & \multicolumn{2}{c}{\textbf{Top-1 RMSD (Å)}} &  \multicolumn{2}{c}{\textbf{Top-5 RMSD (Å)}} \\ 
~ & \% $<2$ & Mean & \% $< 2$ & Mean  \\ \hline
DIFFDOCK (10) & \pmb{$30.53\pm 0.79$} & $7.58\pm 0.32$ & $38.56\pm 1.10$ & $4.81\pm 0.21$  \\ \hline
Autodock Vina (10) & $27.36\pm 0.94$  & $4.83\pm 0.04^*$  &  \pmb{$46.12\pm 1.71$} & $3.40\pm 0.20^*$  \\ \hline 
QMD (10, no local) & $23.58\pm 0.84$ & $5.74\pm 0.11$ & $37.88\pm 0.71$ & $3.59\pm 0.04$ \\ \hline
QMD (10) & $27.93\pm 1.14$ & $5.46\pm 0.04$ & $43.18\pm 1.45$ & $3.28\pm 0.05$  \\ \hline
QMD (10, perturbation) & $29.36\pm 0.81$ & \pmb{$5.38\pm 0.10$} & $45.45\pm 1.98$ & \pmb{$2.92\pm 0.02$}  \\ \hline
\end{tabular}
\end{table*}

We consider the RMSD between the predicted docking pose and the ground-truth (crystal pose). 
The comparison of self-docking is shown in \tref{table:self-docking}.
The top-1 RMSD is the RMSD of the highest ranked pose; the top-5 RMSD is the smallest RMSD among the 5 highest ranked poses. We show the success rate (i.e., percentage of predicted poses with RMSD $<2$Å) and the mean RMSD for top-1 and top-5 respectively.
The experiments are repeated for 3 times and the average success rate with standard deviation is shown.
QMD with perturbation detection improves the top-1 success rate about 2\% compared to Autodock Vina, and improves the top-5 success rate about 6\% compared to DIFFDOCK.
The local continuous search in QMD can improve the top-1 success rate about 4\% based on the results from hSB, and the perturbation detection can improve the success rate about 2\%.

However, the small standard deviation of the mean RMSD of Autodock Vina is because for several samples, Autodock Vina can not output any valid docking result with its embedding selection process. Therefore, the mean RMSD is averaged over obtained results usually with good RMSDs, which is not the mean RMSD for the whole dataset.

The fact that the crystal pose is not guaranteed to be the lowest-energy pose under such a model explains the difficulty to optimize the docking pose. We find that among the 363 testing protein-ligand pairs, about 65\% predicted docking poses in QMD have lower interaction energy than the crystal poses. This implies that our approach has the ability to find the solution with as low as possible objective function, which unfortunately is not the crystal pose. 
The fine construction of the score function may be an independent and important research in the structure-based drug design, but is not the main focus in our study.

Further, compared with the re-docking scenario shown in \tref{table:re-docking}, the higher success rate is due to the rigid-body condition which constrains the number of possible poses. However, in self-docking with the degree of freedom of torsion, it may find other better solutions by virtue of a more suitable configuration, which can not be reached in re-docking.

\section{Conclusions}\label{sec:conclusions}

In this work, we proposed an efficient QMD approach to solve MD problem.
We constructed two binary encoding methods to efficiently discretize the degrees of freedom. 
We determined the interaction energy function as the objective function, and proposed a smoothing filter to rescale the rugged objective function, which significantly benefits further optimization.
We proposed a new hSB algorithm showing great advantage in optimizing over rugged energy landscapes. An adaptive local continuous search is introduced for further optimization of the discretized solution from hSB. 
We also proposed the perturbation detection to help rank the optimized poses, which considers the stability of docking and effectively improves the success rate.
We demonstrated the performance of our QMD approach on the same testing dataset as DIFFDOCK with respect to RMSD and success rate metrics. 
In pocket re-docking scenario, QMD increases the top-1 success rate of the root-mean-square distance (RMSD) less than 2Å by about 6\% compared with DIFFDOCK.
In pocket self-docking scenario, QMD with perturbation detection increases the top-1 success rate by about 2\% compared with Autodock Vina, and increases the top-5 success rate of RMSD less than 5Å by about 6\% compared with DIFFDOCK.

This work indicates that quantum-inspired algorithms have potentials to solve practical problems even before quantum hardware become mature, including molecular configuration conformation, molecular docking and protein folding problems in the drug discovery field.
More practical scenarios have appeared to seek solution possibilities of QAOA, quantum annealing, and quantum computation in the near future.
This work opens a new direction for solving MD problem. 
In the future, it is also worth studying the generalization of QMD in blind docking scenario. Customized parameter settings based on the characteristics of the target ligand and protein can also be considered.

Many technique details including the binary encoding methods, the smoothing filter, and the hSB algorithm in our work have numerous applications in other similar scenarios. The generalization of hSB in other classes of problems is worth further study.

\section*{Acknowledgments}

Y. Li is grateful to S. Pan and J. Hu for insightful discussions.
B. Liu is partially supported by the National Natural Science Foundation of China (Grants No.~12101394 and No.~12171426)


\providecommand{\latin}[1]{#1}
\makeatletter
\providecommand{\doi}
  {\begingroup\let\do\@makeother\dospecials
  \catcode`\{=1 \catcode`\}=2 \doi@aux}
\providecommand{\doi@aux}[1]{\endgroup\texttt{#1}}
\makeatother
\providecommand*\mcitethebibliography{\thebibliography}
\csname @ifundefined\endcsname{endmcitethebibliography}  {\let\endmcitethebibliography\endthebibliography}{}

\newpage

\begin{figure}[ht]
\centering
\includegraphics[width=\textwidth]{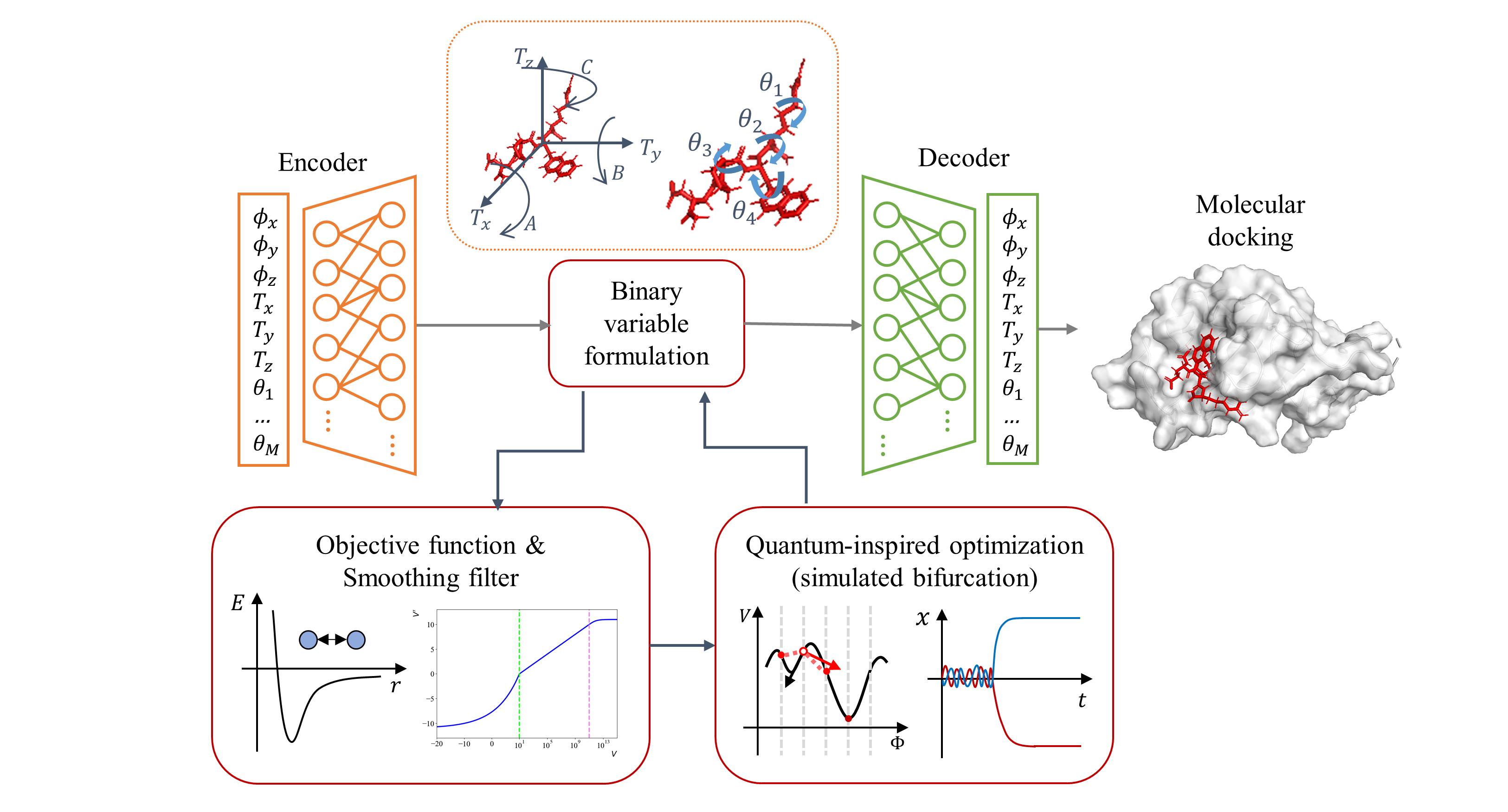}
\caption{For Table of Contents Only}
\label{fig:TOC}
\end{figure}

\end{document}


\title{Quantum molecular docking with quantum-inspired algorithm: Supplemental Material}
	\maketitle

\section{Torsion update}\label{app:torsion update}


Rotatable bonds are bonds that can split the molecule into two nonempty disjoint fragments. The whole molecular structure can be split into several disjoint fragments. The number of rotatable bonds depends on the specific structure of the considering ligand. However, we can set the upper bound of the number of rotatable bonds and select according to the rank of betweenness centrality.

We apply the $M$ torsion updates in a given order of rotatable bonds. For each torsion update $k$, the torsion is applied on the fragment with less number of atoms. The torsion update $k$ of atom $i$ in the target fragment is as follows:
\begin{align}\label{eq:rotation one}
\mathbf{R}_k(\vec{r_i}) = \tilde{R}(\theta_k ) (\vec{r}_i - \vec{r}_{k0}) + \vec{r}_{k0},
\end{align}
where $\theta_k$ is the corresponding torsion angle of the $k$th rotatable bond, and $\vec{r}_{k0}$ is the position of one fixed atom on the corresponding rotatable bond.
The rotation operator $\tilde{R}$ along a rotatable bond of angle $\theta$ reads
\begin{align}
\tilde{R}(\theta)=I +\sin \theta K + (1-\cos \theta) K^2,
\end{align}
where $K$ depends on the unit vector along the rotation axis $\vec{k}=(k_x, k_y,k_z)$
\begin{align}
K=
\begin{pmatrix}
0 & -k_z & k_y \\
k_z & 0 & -k_x \\
-k_y & k_x & 0
\end{pmatrix}.
\end{align}

\section{Interaction energy function}\label{app:energy function}

In this section, we show the details in LJ potential and electrostatic potential which are determined here https://github.com/ccsb-scripps/AutoDock-Vina.

\subsection{Details in LJ potential}

Before calculation the LJ potential, the distance $r_{ij}$ between atom $i$ and $j$ is smoothened according to whether these two atom spheres are overlapped or too far way as follows:
\begin{equation}
r_{ij} \to 
\begin{cases}
r_{ij} - \alpha_s, \quad &\rmif \ r_{ij} > R_{ij} + \alpha_s, \\
r_{ij} + \alpha_s, \quad &\rmif \ r_{ij} < R_{ij} - \alpha_s, \\
r_{ij}, \quad &\rmif \  R_{ij} - \alpha_s < r_{ij} < R_{ij} + \alpha_s,
\end{cases}
\end{equation}
where $R_{ij}$ is the sum of radius of atom $i$ and $j$, and $\alpha_s$ is a smoothing constant usually set to 0.25.

The depth of the potential well of atom $i$ and $j$ is determined as follows:
\begin{align}
\varepsilon_{ij} = \sqrt{\varepsilon_{i}\varepsilon_{j}},
\end{align}
where $\varepsilon_{i}, \varepsilon_{j}$ is the depth of the potential well of each atom, whose value depends on the atom type.

The values of radius and depth of the potential well of the element types included in the testing dataset are shown in \tref{table:para}.

In addition, $V_{\mathrm{LJ}}^{ij}$ is upper bounded by $V_{\max}=100000$ and $V_{\mathrm{LJ}}^{ij}=0$ if $r_{ij} > r_{\max}=8$Å.
The weight of LJ potential is given as $w_1=0.1406$.

\begin{table}[h]
\caption{Values of radius and depth of the potential well of the element types included in the testing dataset.}
\label{table:para}
\centering
\begin{tabular}{c|c|c}
\toprule
\textbf{atom type} & \textbf{radius} & \textbf{\makecell[c]{depth of \\ potential well $\varepsilon$}} \\ \hline 
\text{C} & 2.000 & 0.150 \\ \hline
\text{O} & 1.600 & 0.200 \\ \hline
\text{N} & 1.750 & 0.160 \\ \hline
\text{S} & 2.000 & 0.200 \\ \hline
\text{H} & 1.000 & 0.020 \\ \hline
\text{P} & 2.100 & 0.200 \\ \hline 
\text{I} & 2.360 & 0.550 \\ \hline
\text{F} & 1.545 & 0.080 \\ \hline
\text{Br} & 2.165 & 0.389 \\ \hline
\text{Cl} & 2.045 & 0.276 \\ \hline
\text{Fe} & 0.650 & 0.010 \\ \hline
\end{tabular} 
\end{table}

\subsection{Details in electrostatic potential}

The correction function $f(r_{ij})$ of $r_{ij}$ is as follows:
\begin{align}
f(r_{ij}) = \frac{1}{332}(-8.5525 + \frac{86.9525}{1+7.7839  \rme^{-0.3153 r_{ij}}}).
\end{align}
In addition, $1/r_{ij}$ is upper bounded by $100000$ and ${V_e^{ij}=0}$ if $r_{ij} > r_{\max}=20.48$Å.
The weight of electrostatic potential is given as $w_2=0.1662$.

\section{Implementation details of docking approaches}\label{app:implementation}


\subsection{Implementation details}\label{app:implementation self-docking}

In re-docking scernario, the conformation is fixed as the crystal structure.
In self-docking scenario, the initial conformation is given by RDKit. 

In QMD, the search space is a $6 \times 6 \times 6 $~Å$^3$ box centered at the initial location of the ligand. The initial locations are randomly generated under the uniform distribution in the $6 \times 6 \times 6 $~Å$^3$ box centered at the ground-truth. Hence the whole search space for QMD is a $12 \times 12 \times 12 $~Å$^3$ box. The initial rotation angles of the ligand are randomly generated by the uniform distribution over the rotation range $[-\pi, \pi)$. The number of rotatable bonds is set at most three. The search range of the torsion angle is set $[-\pi/16, \pi/16)$ since the average RMSD between the crystal conformation and the conformation optimized by RDKit of the testing dataset is only 0.85~Å. 
The degrees of freedom of translation and rotation are encoded by four binary variables each and the degree of freedom of torsion is encoded by two binary variables. Thus the step sizes of translation, rotation and torsion are $\Delta T = 0.375$~Å, $\Delta \phi=22.5^{\circ}$, and $\Delta \theta = 5.625^{\circ}$. 
In perturbation detection, we set the small difference about $\delta=0.5\%$. For those within $\delta$ difference of the lowest energy, the best pose is the one with the smallest standard deviation.
In general, to realize larger search space in pocket docking, we can divide it into smaller boxes and run our approach in parallel for each box with several random initial poses.

In Autodock Vina, the search space is a $12 \times 12 \times 12 $~Å$^3$ box centered at the ground-truth. 
The initial location is generated by a random translation from the location of the ground-truth, and the initial rotation angles of the ligand are randomly generated under the uniform distribution over the rotation range $[-\pi, \pi)$. The parameter "exhaustiveness" is set to be 10, which means randomly sample 10 initial pose for each task.

In DIFFDOCK,  the initial location is generated by a random translation from the location of the ground-truth under a uniform distribution $[-6,6]$~Å.

\subsection{Search space}

Here we make some remarks of the search space in Autodock Vina. The choice of search space strongly effects the behaviour of Autodock Vina. 

We find that Autodock Vina shows better success rate if the center of the search space is closer to the ground-truth, which means this method prefers to search the center of the search space. 
Suppose the search space is a $20 \times 20 \times 20 $~Å$^3$ box. 
When we randomly move the search space center away from the ground-truth within 2, 5, 7.5 and 10Å in three dimensions, the top-1 success rates are 31.75\%, 27.02\%, 20.61\% and 10.58\%, respectively. This implies that even though the ground-truth is in the search space, Autodock Vina can not succeed if the ground-truth is in the corner or far away from the center.

Autodock Vina restricts the whole ligand structure to be inside the search space. The default search space in Autodock Vina, SMINA and GNINA is an automatic box around the ligand with the default buffer of $2$Å on all 6 sides. 
Under this setting, the top-1 success rate is about 7\% higher than the setting of $12 \times 12 \times 12 $~Å$^3$ search box centered at the ground-truth.
This setting has already utilized the information of the ground-truth pose and will prefer the pose near the ground-truth. Otherwise, if the ligand is in a totally different orientation and far away from the center, the whole ligand structure is partly outside the search box, which may be forbidden in Autodock Vina. Therefore, the ground-truth is more likely to be found because of the restriction of the search space.

However, in our QMD, we do not have such restriction. The search space is only a restricted space for the center of the ligand instead of the whole ligand structure. QMD allows more possibilities of poses to be explored.

In this work, to compare with Autodock Vina as fair as possible, we set the search space to be $12 \times 12 \times 12 $~Å$^3$ box centered at the ground-truth. This is a more general and natural setting in the real docking experiments.

\section{Machines and runtime details of docking approaches}\label{app:runtime}

The average runtime and devices used are shown in \tref{table:runtime}.
In DIFFDOCK, the final score model was trained on four 48GB RTX A6000 GPUs for 850 epochs around 18 days, and the confidence model was trained on a single 48GB GPU \cite{corso2023diffdock}.
We directly utilize the trained model.
The inference time of DIFFDOCK is calculated by us on a NVIDIA A100-SXM-80GB GPU when generating 10 samples. 
Autodock Vina (realized by ODDT) and QMD are run on the platform for running based on NUMA nodes, featuring 32 Intel(R) Xeon(R) CPUs E7-8890 v4 @ 2.20GHz.
The runtime of Autodock Vina and QMD is the time for per prediction which runs on one CPU when generating 10 samples.
However, the Autodock Vina engine within ODDT is run by C++ compared with QMD by Python.
QMD does not need any pre-training or pre-calculating. The runtime of QMD is the whole end-to-end calculation time, which can be furthered reduced if consider further optimization to run a GPU.

Additionally, in QMD all variables can be simultaneously updated at each iteration step if parallel computation is used. But for the current computational cost shown in  \tref{table:runtime}, the difference defined in Eq.~(15) in the main text is calculated one by one and then update all variables.
Therefore the time cost of QMD can be theoretically reduced by the times of the number of variables if parallel computation is used to calculate the difference. 


\begin{table}[t]
\caption{Average runtime for Autodock Vina, DIFFDOCK and QMD.}
\label{table:runtime}
\centering
\renewcommand{\arraystretch}{1.5}
\begin{tabular}{p{4cm}<{\centering} |p{3.6cm}<{\centering} | p{1.5cm}<{\centering}  }  
\toprule
\textbf{Methods} & \textbf{Avg. time} & \textbf{Device} \\ \hline
Autodock Vina  &  2033 ($\pm5503$) s & 1-CPU \\ \hline
DIFFDOCK (train) &  18 days & 4-GPU \\ \hline
DIFFDOCK (inference) &  10 s & 1-GPU \\ \hline
QMD  &  4468 ($\pm3451$) s & 1-CPU \\ \hline
\end{tabular}
\end{table}

\section{Self-docking visualization}\label{app:visual}

In this section, we show two examples of the self-docking predictions from QMD. The complexes are 6oxy and 6g2b.
We show the RMSDs and scores of the top 5 predictions in \tref{table:6oxy} and \tref{table:6g2b} for 6oxy and 6g2b, respectively. Recall that the score function is the same as interaction energy function. The higher ranked prediction has lower score.

For 6oxy, the score of the ground-truth is -4.21. The higher ranked prediction basically has smaller RMSD, which is effective to find the best prediction close to the ground-truth. However, the top-1 prediction still has lower score than the ground-truth.

\begin{table}[ht]
\caption{The RMSDs and scores of the top-5 predictions of 6oxy.}
\label{table:6oxy}
\centering
\renewcommand{\arraystretch}{1.5}
\begin{tabular}{c|c|c}
\toprule
\textbf{Rank} & \textbf{RMSD} & \textbf{Score} \\ \hline
1 & 0.89 & -4.44 \\ \hline
2 & 0.91 & -4.43 \\ \hline
3 & 5.51 & -3.80 \\ \hline
4 & 9.57 & -3.64 \\ \hline
5 & 4.93 & -3.34 \\ \hline
\end{tabular}
\end{table}

For 6g2b, the score of the ground-truth is -5.52. The top 2 predictions have the lowest scores but the corresponding RMSDs are larger than the other predictions. Therefore, the top-1 prediction is also reasonable and worth testing in the forward experiments. 
The top 5 predictions include the near-ground-truth prediction.

\begin{table}[ht]
\caption{The RMSDs and scores of the top-5 predictions of 6g2b.}
\label{table:6g2b}
\centering
\renewcommand{\arraystretch}{1.5}
\begin{tabular}{c|c|c}
\toprule
\textbf{Rank} & \textbf{RMSD} & \textbf{Score} \\ \hline
1 & 4.77 & -6.16 \\ \hline
2 & 4.37 & -5.87 \\ \hline
3 & 0.54 & -5.76 \\ \hline
4 & 0.60 & -5.75 \\ \hline
5 & 0.47 & -5.69 \\ \hline
\end{tabular}
\end{table}

\begin{figure*}[ht]
\centering
\subfigure[6oxy.]{
\begin{minipage}{9cm}
\centering
\includegraphics[scale=0.39]{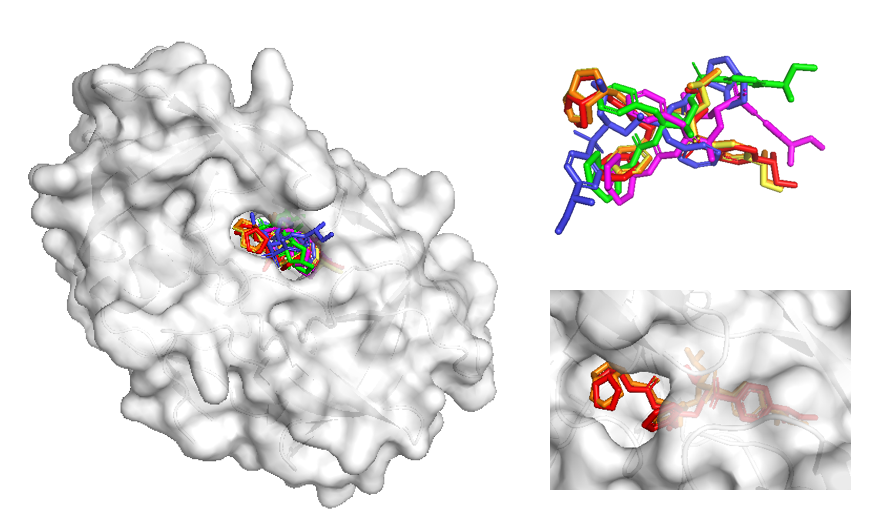}
\label{fig:6oxy}
\end{minipage}}
\subfigure[6g2b.]{
\begin{minipage}{8cm}
\centering
\includegraphics[scale=0.35]{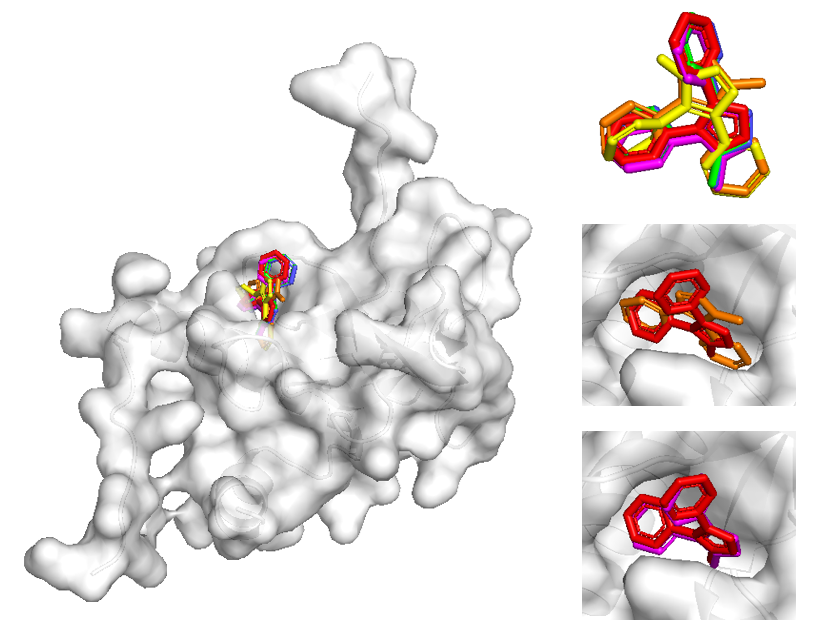}
\label{fig:6g2b}
\end{minipage}}
\caption{Two examples of self-docking. The ground-truth pose is colored red, and the top 5 predictions are colored orange, yellow, green, blue and magenta, in the order of the rank. Shown are the predictions with and without the protein. The top-1 prediction (orange) and top-5 prediction are shown with the protein. 
(a) 6oxy. Top-1 and top-5 predictions are the same (orange).
(b) 6g2b. Top-5 prediction is ranked no.5 (magenta).}
\label{fig:visual}
\end{figure*}

QMD is capable to find a ranked pool of hypotheses of favorable-scoring predictions which are valid and reasonable as shown in \fref{fig:visual}. The ground-truth pose is colored red, and the top 5 predictions are colored orange, yellow, green, blue and magenta, in the order of the rank. We show the predictions with and without the protein. The top-1 prediction (orange) and top-5 prediction (i.e., the prediction with the smallest RMSD among the top 5) are shown with the protein. For 6oxy, top-1 and top-5 predictions are the same; for 6g2b, top-5 prediction is ranked no.5 (magenta).

\bibliography{docking_ref}